\documentclass{emulateapj}

\begin{document}

\title{Anisotropic Galactic Outflows and Enrichment of the Intergalactic 
Medium. II. Numerical Simulations}

\author{Steeve Pinsonneault,\altaffilmark{1,2}
Hugo Martel,\altaffilmark{1,2} and
Matthew M. Pieri\altaffilmark{3}}

\altaffiltext{1}{D\'epartement de physique, de g\'enie physique et d'optique,
Universit\'e Laval, Qu\'ebec, QC, G1K 7P4, Canada}

\altaffiltext{2}{Centre de Recherche en Astrophysique du Qu\'ebec}

\altaffiltext{3}{Department of Astronomy, Ohio State University,
Columbus, OH, 43210}

\begin{abstract}

We combine an analytic model for anisotropic outflows and galaxy formation 
with numerical
simulations of large-scale structure and halo formation to
study the impact of galactic outflows
on the evolution of the Intergalactic
medium. Using this algorithm, we have simulated the evolution
of a comoving volume of size $(15\,{\rm Mpc})^3$ in the $\Lambda$CDM
universe. Using an N-body simulation starting at redshift $z=24$,
we follow the formation of $20\,000-60\,000$ 
galaxies and simulate the galactic outflows produced by these galaxies,
for five outflow opening angles, $\alpha=60^\circ$, $90^\circ$,
$120^\circ$, $150^\circ$, and $180^\circ$ (isotropic outflows).
Anisotropic outflows follow the path of least resistance and thus
travel preferentially into low-density regions, away from cosmological
structures (filaments and pancakes) where galaxies form. These
anisotropic outflows are less likely to overlap with one another
than isotropic ones. They are also less likely to hit pre-galactic collapsing 
halos and strip them of their gas, preventing a galaxy from forming. 
Going from $180^\circ$
to $60^\circ$, the number of galaxies that actually form doubles, 
producing twice as many outflows, and these outflows overlap 
to a lesser extent.
As a result, the metal
volume filling factor of the IGM goes from 8\% for isotropic outflows
up to 28\% for anisotropic ones.
High-density regions are more efficiently enriched  than low density ones 
($\sim80\%$ compared to $\sim20\%$ by volume), even though
most enriched regions are low-densities.
Increasing the anisotropy of outflows increases the extent of
enrichment at all densities, low and high. 
This is in part because anisotropic outflows are more numerous.
When this effect is factored-out, we find that
the probability a galaxy will enrich systems at densities up to $10\bar\rho$ is
higher for increasingly anisotropic outflows. This is interpreted as
an effect of the dynamical evolution of the IGM. Anisotropic outflows 
expand preferentially into underdense gas, 
but that gas can later accrete onto overdense structures.
 The inclusion of photoionization
suppression of low-mass galaxy formation reduces the 
degree of late galaxy formation
and preferentially suppresses galaxy formation in 
low-density regions. The result 
is a decline in the physical extent of galactic 
outflows after $z=3$ as accretion
overwhelms the expansion of new outflows and reduces feedback in 
underdense regions.

\end{abstract}


\section{INTRODUCTION}

The evolution of the Intergalactic medium (IGM) can be significantly 
affected by feedback effects from galaxies. Supernovae and active galactic
nuclei (AGN) can deposit large amounts of energy into the surrounding 
interstellar gas, accelerating it to large velocities. 
If the gas becomes unbound, a galactic wind, or outflow, will result,
with important consequences for the evolution of the IGM.
Galactic outflows deposit energy, momentum, and metal-enriched gas into the
IGM. This can affect the subsequent formation of other galaxies,
by striping the gas from collapsing halos, reheating and
possibly ionizing the IGM, and modifying its cooling rate through metal
enrichment. These metals have been observed via the 
Lyman-$\alpha$ forest
(e.g. \citealt{my87,schayeetal03, ph04,p09,a08,p10}).
Feedback from galactic outflows have been invoked to
explain several observations at the galaxy scale, such as the
high mass-to-light ratio of dwarf galaxies, the overcooling problem, 
and the size of the galactic discs (the angular momentum problem). 
At larger scales, feedback is needed to explain the observed metallicity
and entropy content of the IGM, and the scaling relations for clusters. 

SNe-driven outflows and AGN-driven outflows are complementary. 
The number of SNe increases with galactic mass, but the depth of the 
potential well that
the gas must climb to escape also increases with galactic mass.
Hence, outflows produced by high-mass galaxies are not much larger than
outflows produced by dwarf galaxies, and since these dwarf galaxies are much
more numerous than high-mass ones, they account for most of the
IGM enrichment due to SNe.
AGN-driven outflows are produced by massive galaxies, because these galaxies
are the ones that harbor AGNs. In this paper, we focus on SNe-driven
outflows. AGN-driven outflows are presented in two separate papers
\citep{gbm09,bmg10}.

Numerical simulations and observations reveal that galactic outflows
tend to be highly anisotro\-pic. Several authors have performed
SPH simulations of explosions inside isolated galaxies 
\citep{mf99}, or inside protogalaxies embeded in larger
cosmological structures \citep{ms01a,ms01b}. These simulations reveal
that outflows tend to be bipolar, with
the energy and metal-enriched gas being channeled 
along the direction of least resistance. These results are supported by
many observations that suggest that outflows are anisotropic and
even bipolar (e.g. \citealt{bt88,fhk90,sbh98,stricklandetal00,vr02}).
There is also indirect support for anisotropic outflows from
observations of metal enrichment at high redshift:
metals are found in average-density regions
far from known galaxies and a substantial scatter
in metallicity is seen, which does not seem to be purely a consequence of
density or galaxy proximity \citep{psa06}. Anisotropic
outflows travel larger distances than isotropic
ones, and can more easily reach some low-density regions, while leaving 
others effectively pristine.

Several numerical studies of the impact of galactic outflows on the evolution
of the IGM have been performed. The different approaches used fall into
three categories. The first approach consists of describing the growth
of large-scale structure and the formation of galaxies in the universe using
either an N-body simulation or a semi-analytical method
(e.g. \citealt{sb01}, hereafter SB01, \citealt{tms01,bsw05}), and to combine
it with an analytical model for describing the evolution of the outflow
\citep{tse93}. The model used assumes that outflows remain isotropic as
they propagate through the IGM and the surrounding structures.

The second approach uses a smoothed particle hydrodynamics (SPH)
algorithm to simulate both the formation of large-scale structure and
galaxies and the outflows themselves. Outflows are generated by
either depositing additional thermal energy into SPH particles at
the location of galaxies, to represent the energy generated by SNe
(e.g. \citealt{theunsetal02b}), or by creating a shell of SPH
particles around galaxies, and giving to these particles a large outward
velocity component (e.g. \citealt{std01,sh03,od06}). 
Outflows that start isotropic will become anisotropic as they
propagate into a non-uniform external medium around galaxies.
However, unlike the analytical outflow model used in the first approach,
SPH simulations have a limited resolution. If the anisotropy of the
outflow is not caused by the density distribution around the galaxies, but
instead by the structure of the galaxies, the simulations will not be 
able to resolve it properly.

The third approach consists
of identifying galaxies in an output from an SPH simulation and calculating
the propagation of outflows from these galaxies in $N_a$ different directions
\citep{aguirreetal01}.
Since the resistance encountered by the outflows will be direction-dependent,
outflows will start isotropic but then become anisotropic as the distance
traveled by outflows will vary with direction.
In this approach, the matter located in the outflow slows down, but 
keeps expanding radially,
while in the second approach described above, the matter located in
outflows can be redirected in a different direction when it encounters
resistance. But the most important limitation of this third approach is that
it ignores the potential effect of feedback on the formation of galaxies, 
since the outflows are introduced {\it a posteriori} into an already
completed SPH simulation.

To study anisotropic outflows in
a cosmological context, {\it including the effect of feedback},
we have returned to the first approach. The analytical model used 
in these studies
for describing the outflows do not suffer from limited resolution, but
assume isotropy. 
In a previous paper (\citealt{pmg07}, hereafter Paper~I), we
presented a modified version of the outflow model, designed to describe
the evolution of anisotropic outflows. We then
combined this model with the analytical Monte-Carlo method of SB01.
We used a filtered Gaussian density field to determine the location, mass,
and collapse redshift of galaxies. Galaxies produce anisotropic outflows
that propagate along the direction of least resistance. When an outflow hits
a region destined to collapse by $z=2$, it either strips it of its gas,
preventing the formation of a galaxy, or else enriches it in metals,
potentially changing its cooling rate and the time necessary to turn it
into a galaxy. We found that the anisotropy significantly
enhanced the metal enrichment of average-to-low-density regions, as
outflow tend to propagate into low-density regions, away from the
cosmological structures (pancakes and filaments) where galaxies reside.

The Monte Carlo approach used in Paper~I provides a simple
description of the formation and growth of structures in the Universe, 
but it has a few drawbacks. First, the mass spectrum of halos 
is discrete, with only 10 different masses being allowed.
Second, the peculiar velocity of halos is ignored; halos form at
the comoving location of density peaks, and remain there throughout
the course of the simulation. Hence, the clustering of halos and its
potential consequences are poorly described. 
Third, the movement and accretion of gas on large-scales is ignored.
Fourth, the 
treatment of halo destruction
by mergers is quite simplistic. To address these four issues, we replace
the semi-analytical approach of Paper~I by a full N-body simulation
of structure formation.

This paper is set out as follows. In \S2, we describe our numerical
method and analytic outflow algorithm. Results are presented in \S3, with and 
without photoionization suppression of galaxy formation.
Implications are discussed in \S4 and conclusions are presented in \S5.

\section{THE NUMERICAL METHOD}

\subsection{The Cosmological Simulations}

We perform a numerical simulation of dark matter structure
 on cosmological scales
by considering a $\Lambda$CDM model with present
density parameter $\Omega_0=0.268$,
baryon density parameter $\Omega_{b,0}=0.0441$, cosmological
constant $\lambda_0=0.732$, Hubble constant
$H_0=70.4\rm\,km\,s^{-1}Mpc^{-1}$ ($h=0.704$),
primordial tilt
$n_s=0.947$, and CMB temperature $T_{\rm CMB}=2.725$, consistent with the
results of {\sl WMAP3}
\footnote{http://lambda.gsfc.nasa.gov/product/map/dr2/params/lcdm\_all.cfm}
 \citep{spergeletal07}.
We simulate structure formation inside a comoving cubic volume of
size $L_{\rm box}=15{\rm Mpc}$, with periodic boundary conditions.
The simulation was performed with a Particle-Particle/Particle-Mesh algorithm
\citep{he81}, using $384^3$ equal-mass 
particles and a $1024^3$ grid. The total mass in the computational volume is
$M_{\rm tot}=1.244\times10^{14}{\rm M}_\odot$, and the mass per particle
is $M_{\rm part}=2.197\times10^6{\rm M}_\odot$. The length resolution
is $5.86\,{\rm kpc}$ comoving.

\subsection{Halos, Merger Trees, and Galaxy Formation}
\label{galform}

The simulation produces dumps containing the positions and velocities of 
the particles. The dumps are regularly spaced in time by a time interval 
$\Delta t=2.5\times10^7{\rm years}$, which gives 131~dumps
between redshifts $z=24$ and $z=2$. In each dump,
we identify halos using the friends-of-friends algorithm
\citep{defw85}, with a linking length equal to 0.2 times the mean
particle spacing. As in Paper~I, we set the minimum mass of halos
at $M_{\min}=7.61\times10^7M_\odot$, corresponding to 34 particles.
As we found in Paper I and \cite{pm07}, halos below this mass do not form 
galaxies with mature stellar populations either because of long cooling times 
or photoionization heating. We do not consider the impact of mini-halos or
Population III stars here, we assume that it is negligible aside from 
pre-enriching halos to a level where 
mature stellar populations can arise.

By combining the catalogs of halos, we can find the ancestry of each halo.
Consider a particular halo $H_i^n$ found in dump $n$. We identify the 
particles contained in this halo, and then examine the previous dump, $n-1$,
to find where these particles were. If 50\%
or more of the particles located in a halo $H_j^{n-1}$ in dump $n-1$
end up in halo $H_i^n$ in dump $n$, then halo $H_j^{n-1}$ is considered
to be a {\it progenitor} of halo $H_i^n$.
Then, we have three possibilities, depending on the number of progenitors:

\begin{itemize}

\item Case A: Halo $H_i^n$ has no progenitor. This halo was formed
by the monolithic collapse of a region containing field particles, 
clumps (halos containing fewer than 34 particles), 
or pieces extracted from other halos that
amounted to less than 50\% of those halos. In this case, we consider that
a new halo is born, the birth redshift being the redshift corresponding
to dump~$n$. Halo $H_i^n$ is a ``leaf'' in a merger tree.

\item Case B: Halo $H_i^n$ has one progenitor. In this case, the halo
already existed in dump $n-1$. Between the two dumps, it might have grown
by accretion of field particles or small clumps, or might have lost
mass by evaporation.

\item Case C: Halo $H_i^n$ has several progenitors. In this case, 
the merger of the progenitors have formed a new halo, and the 
progenitors no longer exist. There is an exception: 
if one progenitor, say halo $H_k^{n-1}$, provides 90\% or more of
the mass of halo $H_i^n$, we do not consider
this to be a merger. Halos $H_k^{n-1}$ and $H_i^n$ are
actually the same halo, which grew by {\it accreting} 
(not merging with) smaller halos.
In this case, halo $H_i^n$ is not treated as a new halo, but as a
preexisting one.

\end{itemize}

\begin{figure}
\begin{center}
\includegraphics[width=1.\columnwidth]{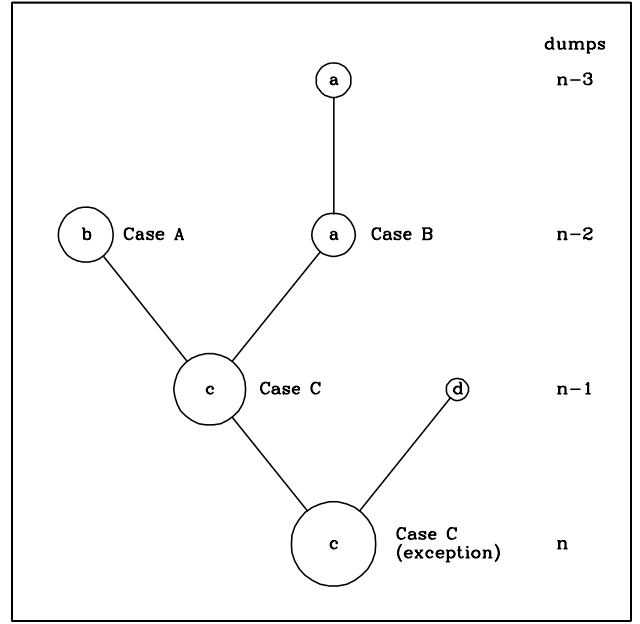}
\caption{Merger tree illustrating the various cases: monolithic collapse
(Case~A), growth (Case~B), merger (Case~C), and accretion (Case~C, exception).
The areas of the circles are proportional to the masses of the halos.
Lowercase labels identify halos, with identical labels indicating halos
that existed in the previous dump.  
}
\label{merger_schem}
\end{center}
\end{figure}

Figure~\ref{merger_schem} illustrates the various cases. Halo b, which had
no progenitor, formed by monolithic collapse (also, halo d, and halo a in
dump $n-3$). Halo a grows by accretion between dumps $n-3$ and $n-2$.
Halo c formed by the merger of halos a and b.
Halo d is accreted onto halo c. This is not regarded as a
merger because 90\% or more of the mass of halo c in dump $n$ came
from one single progenitor.

Once a halo has formed, if its gas is not being kept hot by the UV background, 
it will lose energy by radiative cooling. We implement this photoionization
heating in manner described by the fiducial approach of \citet{pm07}.
This is manifest in the simulation as a redshift-dependent minimum halo mass 
for gas cooling using the mass for which half the halo gas is expected to cool 
(taken from \citealt{d04}). This corresponds to a halo circular velocity of
\begin{equation}
\upsilon_{1/2}=\big[89.4\pm2.4-(6.3\pm0.3)z_{\rm coll}\big] {\rm km\,s^{-1}},
\label{eqv1/2}
\end{equation}
where $z_{\rm coll}$ is the collapse redshift of the halo.
We conservatively apply this 
criterion by requiring that
halos must have reached turnaround after $z=6$ 
for suppression of cooling to occur.
We investigate the impact of photoionization heating 
in \S\ref{results} by showing
results, first without, then with this criterion.

Where photoionization suppression does not occur, halo gas cools 
resulting in galaxy formation and outflow production. As in SB01 and Paper~I,
we calculate the virial temperature of each halo from its mass and
collapse redshift using the method of \citet{ecf96}, and then we
calculate the cooling time using the cooling model of
\citet{wf91}. We assume that the halo turns into a galaxy after
the cooling time has elapsed. As it turns out, for the range of halo masses 
and collapse redshifts considered in our simulations, the cooling times are
almost always very long or very short, that is, either longer than the
age of the universe or shorter than the time interval 
$\Delta t=2.5\times10^7{\rm years}$ between dumps.\footnote{This is caused
by the rapid drop in cooling rate at temperatures $T<10^4{\rm K}$.}
Hence, most collapsed halos either become galaxies almost immediately,
or never become galaxies.
We assume that once a galaxy has formed, the outflow starts immediately.
This is a valid assumption, since the lifespan of the progenitors of 
Type~II SNe
are short compared to the time resolution of our simulations. 

Mergers and accretion of halos onto more massive halos bring an end to the 
fueling of outflows from hosted galaxies, since these galaxies no longer exist 
as independent systems. The effect is to revert any outflows into a 
post-supernova phase (see the following section) and begin the process of 
describing the properties of the new merged galaxy as a source of outflow 
fueling.

\subsection{The Outflows}

\subsubsection{Direction of Least Resistance}

To determine the direction of least resistance out of a halo, we
first determine the orientations and lengths of the semimajor axes of the
halo, using the standard method described by \cite{defw85}. We calculate 
the quadrupole tensor ${\bf Q}$ of the halo,\footnote{Incorrectly called
``inertia tensor'' by many authors.}
\begin{equation}
Q_{ij}=\sum_km_k({\bf r}_k)_i({\bf r}_k)_j\,;\qquad i,j=x,y,z\,;
\end{equation}

\noindent where $m_k$ and ${\bf r}_k$ are the mass of particle $k$
and its position relative to the center of mass, respectively.
We then diagonalize this tensor.
The semimajor axes $a_1$, $a_2$, $a_3$ of the halo are related to
the eigenvalues $Q_1$, $Q_2$, $Q_3$ by $a_i=(5Q_i/M)^{1/2}$,
where $M$ is the mass of the halo, and the eigenvectors give us
the direction of these axes. We then take the direction of least
resistance as being along the shortest axis. 

We have to make sure that the number of particles per
halo is sufficient for this method to work. In Paper~I, the
smallest halos we considered had a mass of $7.61\times10^7M_\odot$.
Here a halo of such mass would be made of 34 particles, which is
sufficient to determine the direction of the shortest axis. 

\begin{figure}
\begin{center}
\includegraphics[width=1\columnwidth]{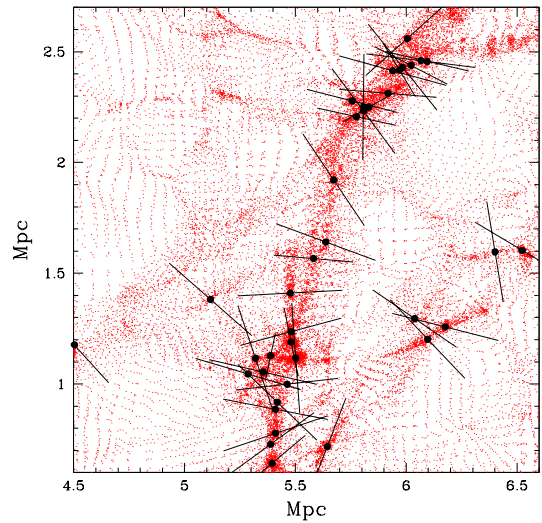}
\caption{Region of size $2.1\,{\rm Mpc}$ and thickness $0.25\,{\rm Mpc}$
centered on
a large-scale filament, at redshift $z=4$.
Red dots: ${\rm P^3M}$ particles;
Black dots: halos. The lines indicate the directions of least resistance.
}
\label{lessres}
\end{center}
\end{figure}

Figure~\ref{lessres} shows a region of size $2.1\,{\rm Mpc}$ and thickness
$0.25\,{\rm Mpc}$ at redshift $z=4$ in the computational volume. We focus
on a region containing a large-scale cosmological filament.
The red dots represent the ${\rm P^3M}$ particles, the black dots indicate the
positions of clusters identified by the friend-of-friend 
algorithm, and the bars indicate
the direction of least resistance. These bars have equal length {\it in 3D},
but since they are seen in projection, they appear to have different lengths.
There is clearly an overall 
tendency of the directions of least resistance to be 
aligned with each other, and to be perpendicular
to the filament. Outflows propagating along these directions will
transport energy and metal-enriched gas preferentially in low-density 
regions.

\subsubsection{Expansion}
\label{expansion}

The dynamical equations describing the evolution of the outflows
are presented in full detail in Paper~I. They are based on the original
treatment of \citet{tse93}, SBO1, and \citet{sfm02}, 
and have been generalized to
the case of anisotropic outflows. Here we present a review of the method.

The injection of thermal energy by SNe
produces an outflow of radius $R$, which consists of a dense shell
of thickness $R\delta$ containing a cavity.
A fraction $1-f_m$, of the gas is piled up in the shell, while a
fraction, $f_m$, of the gas is distributed inside the cavity.
We normally assume that most of the gas is located inside a thin shell
($\delta\ll1$, $f_m\ll1$). This is called the {\it thin-shell approximation}.

The evolution of an outflow of radius $R$ and opening angle $\alpha$
expanding out of a halo
of mass $M$ is described by the following system of equations,
\begin{eqnarray}
\label{rdotdot}
\ddot R&=&{8\pi G(p-p_{\rm ext})\over\Omega_bH^2R}-{3\over R}(\dot R-HR)^2
-{\Omega H^2R\over2}-{GM\over R^2}\,,\\
\label{pdot}
\dot p&=&{L\over2\pi R^3[1-\cos(\alpha/2)]}-{5\dot Rp\over R}\,,
\end{eqnarray}

\noindent where a dot represents a time derivative, $\Omega$, $\Omega_b$,
and $H$ are the total density parameter, baryon density parameter, and
Hubble parameter at time, $t$, respectively.
$L$ is the luminosity (see below), $p$ is the thermal pressure
resulting from this luminosity, and $p_{\rm ext}$ is the external
pressure of the IGM. $\alpha$ is the opening angle of the outflow,
which is $180^\circ$ for isotropic outflows.

The external pressure $p_{\rm ext}$ depends upon the density and
temperature of the IGM. As in Paper~I, we
assume a photoheated IGM made of ionized hydrogen and singly-ionized helium,
with a fixed temperature
$T_{\rm IGM}=10^{4}{\rm K}$ \citep{mfr01}, and
an IGM density $\rho^{\phantom1}_{\rm IGM}$ equal to the mean baryon density,
$\bar\rho_b$. The external pressure is then given by
\begin{equation}
\label{pext}
p_{\rm ext}(z)={\bar\rho_bkT_{\rm IGM}\over\mu}=
{3\Omega_{b,0}H_0^2kT_{\rm IGM}(1+z)^3\over8\pi G\mu}\,,
\end{equation}

\noindent where $z$ is the redshift and $\mu=0.611$ is the mean 
molecular mass.

The luminosity $L$ is the rate of energy deposition or 
dissipation within the outflow, and is given by
\begin{equation}
L(t)=L_{\rm SN}-L_{\rm comp}\,,
\end{equation}

\noindent where $L_{\rm SN}$ is the total luminosity of the supernovae 
responsible for generating the outflow, and $L_{\rm comp}$ represents
cooling due to Compton drag against CMB photons 
(other contributions are neglected, as in Paper~I).
The supernovae luminosity, for a galaxy forming in a halo 
of mass $M$, is given by
\begin{equation}
L_{\rm SN}={f_{\rm w}E_0 \over t_{\rm burst}} {M_*\over M_{\rm req}}
=2.86f_{\rm w}f_*\left({\Omega_{b,0}\over\Omega_0}\right) 
\left({M\over1{\rm M}_\odot}\right)\,{\rm L}_\odot\,,
\label{lum}
\end{equation}

\noindent where $f_*$ is the star formation efficiency,
$M_*=f_*M\Omega_{b,0}/\Omega_0$ is the total mass in stars formed during 
the starburst, $f_{\rm w}$  is the fraction of the total energy
released that goes into the outflow,
$M_{\rm req}$ is the mass of stars
required to form one SN, and $E_0=10^{51} {\rm ergs}$ is
the energy released by each of these SNe.
As in Paper~I, we take the values $f_*=0.1$,
$M_{\rm req}=89.7 {\rm M_\odot}$ (derived using a 
broken power-law IMF from \citealt{k01}),
and the mass-dependent expression for $f_{\rm w}$ given by \citet{sfm02}.

The Compton luminosity for a mixture of ionized hydrogen and singly-ionized 
helium is given by
\begin{equation}
\label{lcomp2}
L_{\rm comp}={2\pi^3\over45}{\sigma_t\hbar\over m_e}
\left({kT_{\gamma0}\over\hbar c}\right)^4
\left(1-\cos{\alpha\over2}\right)(1+z)^4pR^3\,,
\end{equation}

\noindent where $\sigma_t$ is the Thompson cross section, and
$T_{\gamma0}$ is the present CMB temperature. We used 
$T_{\gamma}=T_{\gamma0}(1+z)$, which is valid over the range
of redshifts we consider.

The expansion of the outflow is initially driven by the supernovae luminosity.
After a time $t_{\rm burst}=5\times10^7{\rm yr}$,
the supernovae turn off, and the outflow enters the ``post-supernova phase.''
The pressure inside the outflow keeps driving the expansion, but this
pressure drops since there is no energy input from supernovae.
Eventually, the pressure will drop down to the level of the external IGM 
pressure. At that point, we assume that
the expansion of the outflow will simply follow 
Hubble expansion.

\subsubsection{Ram Pressure Stripping}

As outflows propagate into the IGM, the expanding shells
of swept-up gas may eventually hit halos. Following
SB01 and Paper~I, we assume that if an already formed galaxy is
hit by an outflow, the cross section is too small for the impact to have
any significant effect. If a halo that has not yet collapsed and formed a
galaxy is hit, two things might happen. Either the ram pressure of the
outflow will strip the halo of its baryonic content, preventing the
formation of a galaxy, or else the halo will be enriched in metals by the 
outflow. The condition for stripping is
\begin{equation}
\left({l^2 \over 4R^2}\right) M_ov_o \ge M_b v_{\rm esc},
\label{strip}
\end{equation}

\noindent
where $M_o$ and $R$ the mass and radius of the shell, respectively,
$v_o$ is the outflow velocity, $l$ and $M_b$ are the radius and baryonic 
mass of the halo being hit, and $v_{\rm esc}$ is the escape velocity from 
the halo. In Paper~I, we used the spherical collapse model to estimate
the radius $l$ at the time of the hit. This model predicts that halos
expand, eventually turn around, and collapse to a point. 
This was straight forward in the Monte Carlo approach of 
Paper~I. However, in the numerical approach used in this paper, we identify
collapsed halos using the friend-of-friend algorithm, and these halos 
obviously have a finite size. Thus, the basic spherical model needs to
be modified to account for the fact that real halos do not collapse down 
to a point, but instead contract by a factor of order 2 after turnaround,
and reach virial equilibrium. We describe our modified spherical
collapse model in Appendix~B.

When the criterion for stripping is met, the halo is labeled as being 
stripped, and it will not form a galaxy.
If the halo is not stripped, the outflow will deposit metals
into the halo.

\subsubsection{Metal Enrichment of Halos}

We assume that halos start with metallicity
of $\rm[Fe/H]=-3$, which is negligible for the purposes of
calculating the halo cooling time.
Once galaxies are formed, metals are produced at rate of
$2{\rm M_\odot}$ per SN \citep{ns98}. Hence the mass of metals in the
outflow is
\begin{equation}
M_Z=f_{\rm esc}{2{\rm M_\odot}\over M_{\rm req}}f_*{\Omega_{b,0}\over\Omega_0}
M\,,
\end{equation}

\noindent
where $f_{\rm esc}$ is the fraction of ISM gas
blown out with the outflow. We use the
value $f_{\rm esc}=0.5$ taken from the numerical simulations of \cite{mfm02}.
This mass of metals is distributed evenly throughout the volume of the outflow.

When an outflow strikes a halo and does not strip it, it modifies 
the metal content of the halo by depositing a fraction of its metals,
$f_{\rm dep}V_{\rm overlap}/V_{\rm outflow}$, where
$f_{\rm dep}$ is a mass deposition
efficiency, which we set at $f_{\rm dep}=0.9$, and
$V_{\rm overlap}$ is the volume of the outflow that overlaps with the halo.
For details, we refer the reader to Paper~I.

The radiative cooling rate increases with metallicity. So
when a halo is enriched by metals, we recompute its cooling time, which
determines when this halo will form a galaxy. This is the potential
to ``bring to life'' dead halos. A halo that cannot become a galaxy
because its cooling time exceeds the age of the universe might form a
galaxy after all, if metal enrichment reduces the cooling time.
It turns out to be a rare occurrence. In all the simulations 
we performed only several galaxies
were formed this way, out of $20\,000-60\,000$ galaxies.

\subsection{Metal Enrichment of the IGM}

Stellar evolution in galaxies produces metals, that are then carried
into the IGM by outflows. Of particular interest is the {\it volume filling 
factor}, that is, the fraction of the IGM, by volume, that has been 
enriched in metals by outflows. 
To calculate the volume filling factor,
we cannot simply add up the final volumes of the outflows,
and divide by the volume of the computational box. Intergalactic
gas enriched by outflows will move with time as structures grow, and
therefore regions that were never hit by outflows might end up
containing metals. 

We take this effect into account, by using a dynamic particle enrichment 
scheme that was developed by \citet{gbm09}. By combining the dumps produced
by the $\rm P^3M$ code, which contains the position of the particles,
with the positions, orientations, and radii of the outflows, we can
find which particles are being hit by outflows, and at what redshift.
We can then use a Smoothed Particle Hydrodynamics technique to estimate
the volume filling factor at any redshift.
A smoothing length $h$ is ascribed to each particle. 
We calculate $h$ iteratively by requiring that each particle 
has between $60$ and $100$ neighbors within a distance $1.7h$.
We then treat each particle as an extended sphere of radius $1.7h$ 
over which it is considered to be spread.
We then divide the computational
volume into $N=256^3$ cubic cells. 
The cells that are covered by one or more enriched particles are then
considered enriched.
Notice that simply counting the number of enriched particles in 
each cell would fail in low-density regions, where the local particle spacing
exceeds the cell size

Additionally,
giving a smoothing length to each $\rm P^3M$ particle enables us to easily 
calculate the gas density in the center of each cell, using the standard 
SPH equation,
\begin{equation}
\label{SPH}
\rho({\bf r}_g)=\sum_im_iW({\bf r}_g-{\bf r}_i,h_i)\,,
\end{equation}

\noindent where ${\bf r}_i$, $m_i$, and $h_i$ are the position, mass, and
smoothing length of particle $i$, ${\bf r}_g$ is the position of a grid point,
and $W$ is the smoothing kernel (see \citealt{monaghan92} for a detailed
description of SPH). We will use this in \S~3.2 below.

\section{RESULTS}
\label{results}

We performed a series of ten simulations:
five with our galaxy formation 
scheme excluding suppression by photoionization heating, 
to produce results that can be compared with our results in Paper I,
and a further five with this suppression
as used in \citet{pm07}. In each case we simulate outflows 
with opening angles
$\alpha=60^\circ$, $90^\circ$, $120^\circ$, $150^\circ$,
and $180^\circ$. This latter case corresponds to isotropic outflows.

\subsection{Without Photoionization Suppresion}

\subsubsection{Volume Filling Factor}

\begin{figure}
\includegraphics[width=1\columnwidth]{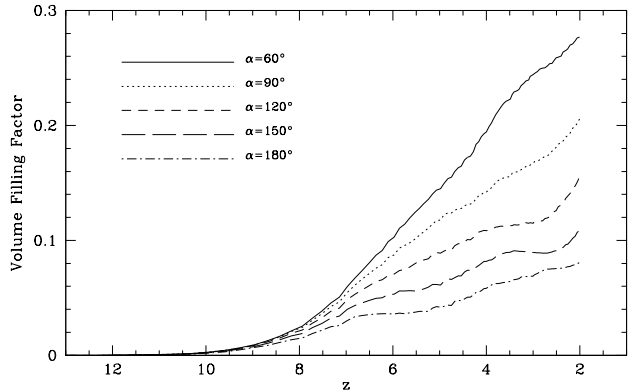}
\caption{Volume filling factor versus redshift,
with various opening angles, for simulations without
photoionization suppression of galaxy formation. 
}
\label{filling_NR}
\end{figure}

Figure~\ref{filling_NR} shows the evolution of the volume filling factor
for the five opening angles considered. Enrichment starts at redshift
$z\sim12$ when the first galaxies form, and steadily increase with time,
to reach $8-28\%$ by redshift $z=2$. The fraction increases with
decreasing angle, at all redshifts. 
This differs from the results of Paper I, and clearly shows the
importance of an accurate description of clustering. In Paper~I, 
we found that the fraction
was nearly constant, near 14\%, for angles between $\alpha=180^\circ$
and $\alpha=100^\circ$, and then dropped down to $11\%$ at 
$\alpha=60^\circ$ (Fig.~8, bottom left panel, in Paper~I). 

Three distinct effects determined the overall volume filling factor 
in Paper I. 
These effects are still present here but their relative impact is changed. 
First, outflows are much more likely to overlap when the sources are highly
clustered and this effect is diminished where outflows are anisotropic 
and aligned. 
Second, clustering makes it more likely that an expanding
outflow will hit a pre-galactic collapsing halo, possibly resulting in
gas stripping and preventing a galaxy from forming and producing its own 
outflow. Again the impact of this effect is diminished where outflows are
 anisotropic. 
Third, the volume of each outflow decreases with 
decreasing angle, the increase in radius not being enough to compensate
for the smaller angle. The first two effects tend to increase the 
volume filling factor for anisotropic outflows and are sensitive to the 
level of galaxy clustering. The third effect tends to decrease the 
overall volume filling factor and is insensitive to the degree of clustering. 
Our results here show a more accurate and stronger degree of halo 
clustering, thus explaining the more marked impact of the first two 
effects with respect to the third. This also explains the small fractions 
we obtain in the isotropic case here ($8\%$ at $z=2$)
by comparison to the equivalent result in Paper I.

\begin{deluxetable}{rccrrc}
\tablecaption{Statistics of Enrichment and Hits Without 
Photoionization Suppresion}
\tablewidth{0pt}
\tablehead{
\colhead{$\alpha$} &
\colhead{$V_{\rm outflows}$} &
\colhead{V.F.F.} &
\colhead{$N_{\rm hit}$} &
\colhead{$N_{\rm stripped}$} &
\colhead{$N_{\rm outflows}$}
}
\startdata
$60^{\circ}$  & 0.388 & 0.277 & $159\,957$ & $145\,188$ & $63\,015$ \cr
$90^{\circ}$  & 0.381 & 0.206 & $192\,115$ & $168\,365$ & $52\,354$ \cr
$120^{\circ}$ & 0.369 & 0.155 & $213\,846$ & $178\,304$ & $44\,399$ \cr
$150^{\circ}$ & 0.367 & 0.109 & $231\,058$ & $178\,781$ & $38\,707$ \cr
$180^{\circ}$ & 0.369 & 0.081 & $234\,365$ & $165\,353$ & $34\,324$ \cr
\enddata
\label{TabFilling_nore}
\end{deluxetable}

In Table~\ref{TabFilling_nore}, 
we list, for each opening angle, the total
volume of the outflows, $V_{\rm outflows}$
(sum of all the volumes, ignoring overlap, in units of the box volume),
the volume filling factor (V.F.F.) 
the number of halos hit and stripped ($N_{\rm hit}$, $N_{\rm stripped}$,
respectively), and the
number of outflows, $N_{\rm outflows}$, 
all at $z=2$. 
Comparing $V_{\rm outflow}$ with V.F.F. in Table~\ref{TabFilling_nore},
we clearly see that the effect of overlap is mild for small
opening angles, but becomes very important for large ones.
As the opening angle increases,
there are more halos hit, more halos stripped, and as a result
fewer galaxies are formed, 
producing fewer outflows. Going from $\alpha=60^\circ$
to $\alpha=180^\circ$, the number of outflows drops by a factor of 
2. Despite this,
though, their total volumes are comparable (0.388 vs. 0.369) since the volume
per outflow is greater for outflows with larger opening angles.

In addition to the three effects described above we are also able to 
consider a fourth effect given the dynamical nature of our new 
simualtions. The matter enriched with metals moves as the
simulation proceeds, and this motion modifies the volume filling factor. 
An isotropic outflow produced by a galaxy enriches 
matter that is located near 
the galaxy, and this matter is likely to later contract as it accretes over 
the cosmological structure hosting the galaxy. Highly anisotropic outflows 
travel larger distances, and might enrich matter located far from 
cosmological structures, and potentially expanding away from them.

Notice that the
number of galaxies formed is significantly larger than in Paper~I
($\sim20\,000$ in a cosmological volume that was 40\% larger). As we showed
in Paper~I, the Monte Carlo method underestimates the number of galaxies
compared to either an N-body simulation or the Press-Schechter approximation
(see Fig.~4 in Paper~I). This provides additional motivation for performing 
a full cosmological simulation.

\subsubsection{Distribution of Metal-Enriched Gas}

\begin{figure*}
\hskip0.7in
\caption{
Slice (of comoving size $15\,{\rm Mpc}\times15\,{\rm Mpc}$,
and comoving thickness $0.1\rm{ Mpc}$) of the computational volume for
an opening angle $\alpha = 180^{\circ}$,
showing the evolution of the distribution of metals, for
simulations without
photoionization suppression of galaxy formation. The areas shown in green
are enriched. The black dots represent the $\rm P^3M$ particles, and show the
large-scale structures.
}
\label{slice180}
\end{figure*}

\begin{figure*}
\hskip0.7in
\caption{
Same as Figure~\ref{slice180}
for an opening angle $\alpha = 120^{\circ}$.
}
\label{slice120}
\end{figure*}

\begin{figure*}
\hskip0.7in
\caption{
Same as Figures~\ref{slice180} and \ref{slice120}
for an opening angle $\alpha = 60^{\circ}$.
}
\label{slice60}
\end{figure*}

Figures~\ref{slice180}--\ref{slice60} show slices of the computational 
volume at redshifts $z=4.98$, 4.00, 3.00, and 2.00, for opening angles
$\alpha=180^\circ$, $120^\circ$, and $60^\circ$, respectively. At high 
redshift, metals are located in high-density regions where the sources 
are located. As time goes on, they eventually reach lower-density region, 
but never travel far enough to reach the deepest voids. Instead, even at
$z=2$, all metals are found near dense structures like clusters and 
filaments.

\begin{figure}
\includegraphics[width=.97\columnwidth]{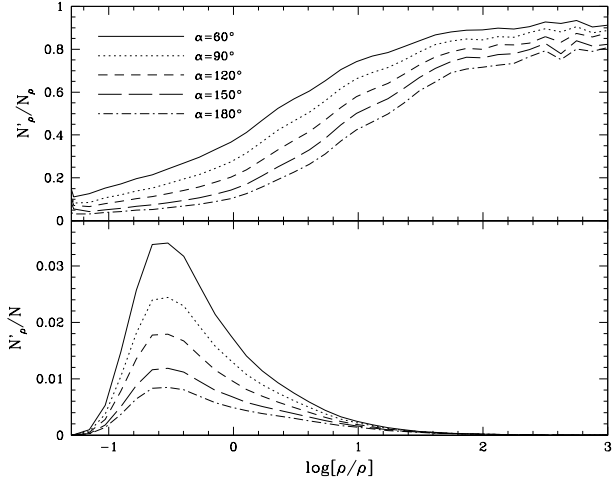}
\caption{The number of enriched cells $N'_\rho$ in the 
simulation volume at $z=2$ 
as a function of IGM density with various opening angles, for
simulations without
photoionization suppression of galaxy formation. {\it Top panel:} 
$N'_\rho$ as a fraction of the number of cells at this density, $N_\rho$.
{\it Bottom panel:} $N'_\rho$ as a function of the total number of cells, 
$N$.
} 
\label{enrich_NR}
\end{figure}

\begin{figure}
\includegraphics[width=.97\columnwidth]{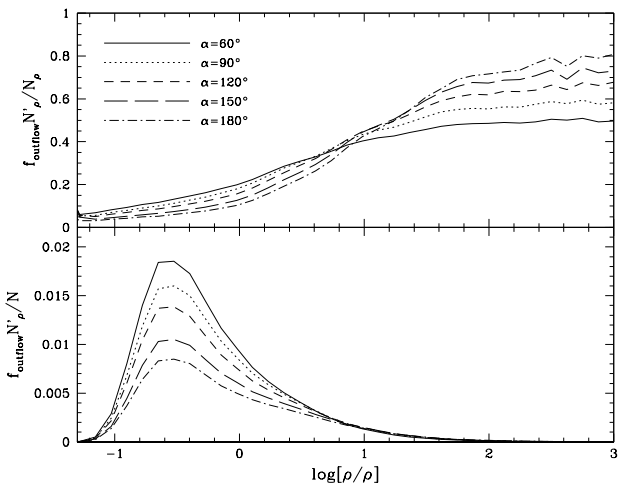}
\caption{Same as Figure~\ref{enrich_NR}, except that the number of
enriched cells has been multiplied by the factor 
$f_{\rm outflow}(\alpha)
\equiv N_{\rm outflow}(180^\circ)/N_{\rm outflows}(\alpha)$,
where $N_{\rm outflows}$ is given in Table~\ref{TabFilling_nore}. 
This isolates the impact of anisotropy on the regions enriched by
individual outflows.}
\label{enrich2_NR}
\end{figure}

The main effect of narrowing the opening angle is to increase 
the volume over which
metals are dispersed, as Figure~\ref{filling_NR} showed. Comparing 
Figures~\ref{slice180}--\ref{slice60}, it appears that both high- and
moderately-low-density regions are affected. It is not obvious from these
figures that anisotropic outflows enrich predominantly low-density
regions. To investigate this issue, we calculated the density at each of the 
$N$ cells on the grid, using equation~(\ref{SPH}). We then
binned these cells according to density, and counted in each bin the
number of cells, $N_\rho$, and the number of enriched cells,
$N'_\rho$. From this, we calculated statistics of metal enrichment
vs. density. The results are shown in Figure~\ref{enrich_NR}.
The top panel shows $N'_\rho/N_\rho$; this is 
effectively the probability of enriching a systems of a given density.
The galaxies producing the outflows are located in high-density regions,
hence these regions are favored over low-density ones, even with anisotropic
outflows (such outflows will still enrich the gas located near the galaxies).
The enrichment reaches 80\% or more in high-density regions.
The extent of enrichment increases with decreasing opening angle, at all
densities, although this effect is most significant at lower densities. 
Going from $\alpha=180^\circ$ to $\alpha=60^\circ$, $N'_\rho/N_\rho$
increases by a factor of 3.90 at $\log[\rho/\bar\rho]=-1$,
3.53 at $\log[\rho/\bar\rho]=0$, 1.75 at $\log[\rho/\bar\rho]=1$,
1.24 at $\log[\rho/\bar\rho]=2$, and 1.13 at $\log[\rho/\bar\rho]=3$.

The bottom panel shows the number of grid points enriched at a given 
overdensity as a fraction of the total number of cells, $N$. 
There are very few cells at densities
below $\log[\rho/\bar\rho]=-1.2$, but about $\sim10\%$ of these cells
get enriched, explaining why $N'_\rho/N$ becomes negligible while
$N'_\rho/N_\rho$ does not. The curves are very similar for all opening
angles, except for the overall amplitude. Most enriched cells are
located at low-densities in the range $\log[\rho/\bar\rho]=-1$ to 0,
even though enrichment is less efficient in these regions compared
to high-density ones. Even in the case of 
isotropic outflows most of metals
are in underdense regions when considered by volume.

As the opening angle is reduced, outflows travel predominantly into
low-density regions, and the number of outflows increases, because less
stripping occurs. These two effects are acting in the same direction
in low-density regions, while they are competing in high-density regions.
To separate the two effects, we replotted in Figure~\ref{enrich2_NR} the same
results as in Figure~\ref{enrich_NR}, but rescaled by the number of outflows.
That is, we have multiplied the number of enriched cells $N'_\rho$ by the
factor $f_{\rm outflow}\equiv N_{\rm outflow}(180^\circ)
/N_{\rm outflows}(\alpha)$, where $N_{\rm outflow}$ can be read from the
last column of Table~\ref{TabFilling_nore}. 
The top panel of Figure~\ref{enrich2_NR}
is strikingly similar to the top panel of Figure~9 in Paper~I. As the opening
angle decreases, low-density regions are being enriched more, at the expense 
of high-density regions. However, the crossover happens at a higher density,
$\rho/\bar\rho\approx 10$, compared to $\rho/\bar\rho\approx 2$ 
if we make a similar correction to $f_{\rm outflow}$ for
Paper~I. We attribute this difference to our dynamical treatment of 
metal enrichment. Enriched gas located in low-density regions can eventually
move into higher-density regions as the system evolves.

\subsection{With Photoionization Suppression}

\begin{deluxetable}{rccrrc}
\tablecaption{Statistics of Enrichment and Hits 
With Photoionization Suppresion}
\tablewidth{0pt}
\tablehead{
\colhead{$\alpha$} &
\colhead{$V_{\rm outflows}$} &
\colhead{V.F.F.} &
\colhead{$N_{\rm hit}$} &
\colhead{$N_{\rm stripped}$} &
\colhead{$N_{\rm outflows}$}
}
\startdata
$60^{\circ}$  & 0.301 & 0.189 &  $84\,599$ &  $80\,082$ & $49\,276$ \cr
$90^{\circ}$  & 0.280 & 0.119 & $107\,242$ & $100\,410$ & $39\,994$ \cr
$120^{\circ}$ & 0.255 & 0.079 & $120\,593$ & $110\,588$ & $32\,773$ \cr
$150^{\circ}$ & 0.238 & 0.065 & $126\,114$ & $112\,347$ & $27\,339$ \cr
$180^{\circ}$ & 0.223 & 0.047 & $120\,717$ & $103\,479$ & $22\,992$ \cr
\enddata
\label{TabFilling_re}
\end{deluxetable}

\begin{figure}
\includegraphics[width=.97\columnwidth]{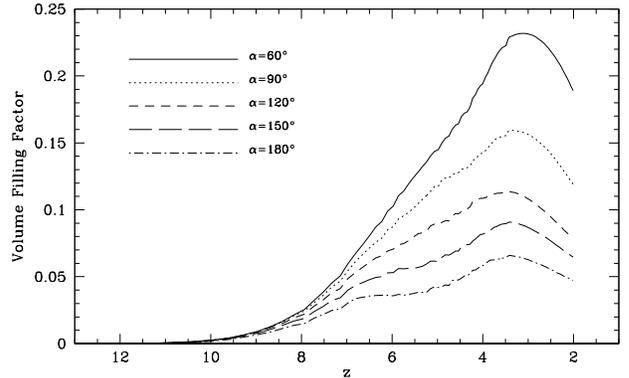}
\caption{Volume filling factor versus redshift,
with various opening angles, for simulations with
photoionization suppression of galaxy formation.}
\label{filling}
\end{figure}

Figure~\ref{filling} shows the evolution of the volume filling factor
with the inclusion of photoionization suppression of galaxy formation 
as described in \S\ref{galform}. By comparison with Figure~\ref{filling_NR}, 
one can see that the volume filling factor remains unaffected until 
$z\approx4$ and after this the extent of mechanical feedback is diminished. 
The volume filling factor continues to increase until $z\approx3$ after 
which {\it it actually begins to fall} until the end of 
our simulations at $z=2$. This is a result of the dynamics included in our
simulation, whereby gravitational attraction carries metal-enriched regions 
towards dense structures faster than outflows can carry them away from 
these structures. The final statistics of the suite of runs with 
photoionization suppression are shown in Table~\ref{TabFilling_re}. The 
same notation as Table~\ref{TabFilling_nore} is used and as before 
increasing opening angles leads to more outflows hitting halos, more 
stripping and so fewer galaxies (and so outflows).

As one can see by comparing Table~\ref{TabFilling_re}
with Table~\ref{TabFilling_nore}, the introduction of photoionization
suppression has the weakest impact for our smallest opening angle: the
volume filling factor is 32\% lower due photoionization heating in the for
$\alpha=60^{\circ}$. However, the most pronounced impact occurs one of
our intermediate models, $\alpha=120^{\circ}$ where the 
volume filling factor is 49\% lower.
The number of halos hit is also non-linear with opening angle, peaking at
$\alpha=150^{\circ}$. It is also notable that the summed volume of all
outflows $V_{\rm outflows}$ shows a clear decline with increasing opening
angle, indicating that the fall in galaxy numbers with increasing opening
angle dominates over the impact of the change in individual outflow volumes.
As in the case without the inclusion photoionization suppression, the true
model volume fraction shows a stronger opening angle dependence and this is
a sign of reduced outflow overlap for more anisotropic outflows.

\begin{figure}
\includegraphics[width=1\columnwidth]{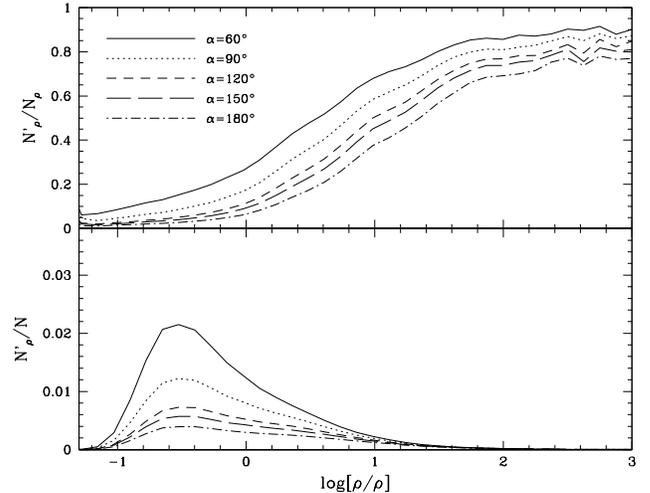}
\caption{The number of enriched cells $N'_\rho$ in the 
simulation volume at $z=2$ 
as a function of IGM density with various opening angles,
for simulations with
photoionization suppression of galaxy formation. {\it Top panel:} 
$N'_\rho$ as a fraction of the number of cells at this density, $N_\rho$.
{\it Bottom panel:} $N'_\rho$ as a function of the total number of cells, 
$N$.
} 
\label{enrich}
\end{figure}

The density dependence of the volume filling factor is shown in
Figure~\ref{enrich}. These results are broadly similar to those for the
suite of simulations 
without photoionization suppression (Figure~\ref{enrich_NR}).
The values of $N'_\rho/N$ are reduced in underdense regions 
($\log[\rho/\bar\rho]<0$), but almost unaffected in very high-density
regions ($\log[\rho/\bar\rho]>1$). In these high-density regions, outflows
strongly overlap, and the removal of some of them by photoionization
suppression does not prevent cells from being enriched.

\section{DISCUSSION}

In Paper~I, we described the evolution of 
anisotropic outflows and the enrichment
of the IGM using a simple Monte Carlo method that was originally introduced
by SB01. In this paper, we used cosmological N-body simulations to simulate
the formation and evolution of large-scale structures in a more realistic
way. The results obtained with both methods show interesting differences.
In the Monte Carlo approach, the
volume filling factors were largely independent of opening 
angle (Fig.~8, bottom-left panel of
Paper~I). Here were find a clear dependence on outflow anisotropy:
the volume filling factor {\it increases\/} 
with decreasing opening angle, as 
Figures~\ref{filling_NR} and~\ref{filling} showed.

In Paper~I, we generated a Gaussian density field at high redshift,
filtered it at various mass scales, and used the spherical collapse
model to predict the location and collapse redshift of halos. Hence,
the location of halos was predetermined by the initial conditions, and
these locations did not change with time. Also, the density $\rho$ was
determined by a nonlinear mapping of the initial density field. 
The density $\rho({\bf r})$ at a comoving position ${\bf r}$ was thus
entirely determined by the initial density at that location. This
approach has the advantage of simplicity, but does not constitute a
full treatment of clustering.

This can have important consequences. Galaxies tend to be located
inside larger cosmological structures, like filaments or pancakes.
If several galaxies are located in a common structure, they
can affect each other in two ways. First, the outflows might overlap, and
second, if an outflow hits a halo that has not yet formed a galaxy,
ram pressure stripping might prevent the galaxy from forming. These two
effects are reduced when the outflows are anisotropic, because
such outflows travel away from the cosmological structures, into
low-density regions, and tend to avoid other galaxies and other
outflows.

These effects were present in the Monte Carlo simulations (Fig.~8, bottom
right panel of Paper~I), however, they were not dominant. Instead they either 
approximately balanced the decline in individual outflow volumes for 
increasingly
anisotropic outflows (down to opening angles of  $\alpha=100^\circ$) or were
dominated by them (opening angles from $\alpha=100^\circ$ to
$\alpha=60^\circ$).
The strength of clustering is much more apparent in the
cosmological N-body simulations, and these effects now dominate. The higher
level of clustering in the simulations here give greater weight to the impact 
of outflows that travel preferentially into low-density regions. Most results
which are sensitive to this effect are all stronger here: the reduced outflow
overlap, the reduced stripping of halos, the increased extent of enrichment
in low-density regions. There is one aspect in which these results show a
weaker signal of anisotropy than Paper~I: the enrichment of overdense systems.
In Paper I the fact that outflows travel into low-density regions resulted in
a reduction in the physical extent of outflows in overdense regions, here 
however, anisotropic outflows increase the extent of outflows for densities 
all the way up to $\rho/\bar\rho\approx 10$. This a 
consequence of the dynamic nature of our N-body simulations. At the time of 
launching the outflow may expand into underdense regions but as the simulation 
progresses the gas enriched by the outflow is 
re-accreted onto overdense structures.

The impact of this accretion is also notable in the context of results from
\citet{o09}. They find that understanding the accretion of outflow gas back
onto galaxies plays an important role in galaxy assembly and the stellar mass
function of galaxies by $z=0$. We find accretion of outflow gas onto
large-scale structures plays an important role at $z>2$ and it seems likely
that much of this gas continues its infall onto galaxies and fuels
star formation.

Consider a galaxy producing an outflow that carries metals into low-density 
regions. As Figures~\ref{slice180}--\ref{slice60} show, outflows never
travel very far from their sources. Thus, the gas enriched by outflows
is likely to be gravitationally bound to the structure hosting the galaxy.
As that gas accretes onto that structure, it carries metals from
a low- to a high-density region. This effect has been found by
\citet{gbm09} in a different context (anisotropic outflows powered by
AGNs). As these authors argue, even though anisotropic outflows
deposit metals predominantly in low-density regions, the subsequent evolution
of the large-scale structure will tend to partly wash out that effect.
The only exception might occur when outflows travel very large distances, 
reaching gas inside deep voids that is not destined
to accrete onto any structure within a Hubble time. This happens with
AGN-powered outflows \citep{gbm09}, but SNe-powered outflows do not
have enough energy to reach such distances.

The simulations presented in this paper use the same outflow
model as the ones presented in Paper~I. Only the method used for generating
the population of halos (Monte Carlo vs. N-body simulations) and for
calculating the enrichment of the IGM (static vs. dynamical) are different.
The outflow model contains several parameters, such as
$f_*$, $f_{\rm w}$, $f_{\rm esc}$, and $M_{\rm req}$. We use the same
values as in Paper~I, and we refer the reader to that paper for
a discussion on the limits on these various parameters.

\section{SUMMARY AND CONCLUSION}

We have combined cosmological N-body simulations with a friend-of-friend 
algorithm for identifying halos and merger events,
and an anisotropic outflow model, to study
the metal-enrichment of the IGM by SNe-powered galactic outflows,
in a $\Lambda$CDM universe. Outflows are modeled as bipolar cones traveling
along the direction of least resistance, which we take as the direction
along which the density drops the fastest. We performed five
simulations with outflow opening angles ranging from $60^\circ$
to $180^\circ$, the latter case corresponding to isotropic outflows.
Each simulation was stopped at redshift $z=2$.
For each simulation, we tracked the deposition of metals in the 
IGM by outflows,
and the subsequent motion of the metal-enriched gas. We also included the
effect of ram pressure stripping and metal enrichment of 
pre-galactic collapsing
halos that are hit by outflows. We did this with and without the 
inclusion of photoionization suppression of galaxy formation resulting in
a total of ten simulations.
Our main results are the following.

\begin{itemize}

\item
Anisotropic outflows travel predominantly into low-density regions,
away from cosmological structures (filament and pancakes) that host
most galaxies. As a result, anisotropic outflows are less likely to
overlap than isotropic ones, and are less likely to inhibit galaxy
formation by stripping collapsing halos of their gas. The combination of
these effect results in an increase of the volume filling factor of the IGM
with decreasing opening angle, from 8\% with isotropic outflows, up to
28\% with opening angle of $60^\circ$.

\item
The majority of collapsing halos hit by outflows end up being 
stripped of their gas, preventing the formation of a galaxies. The total
number of galaxies formed drops by a factor of 2 when going from
opening angle $\alpha=60^\circ$ to $180^\circ$. Halos not stripped
are enriched in metals, with negligible consequences for the formation
time of the galaxy. In particular, fewer than one in $10^3-10^4$ galaxies
that formed would not have formed without the hit, because their
cooling time exceeded the Hubble time.

\item
The volume filling factor remain small (28\% or less) because
outflows do not travel large distances. They
remain fairly close to the structures where they originate, never
reaching the center of the deep cosmological voids. The regions
most efficiently enriched (up to 90\% by volume) are the high-density
regions, where the galaxies producing the outflows are located.
The enrichment of low-density regions is not as efficient. However,
by volume, most metals are located in these low-density region,
because their combined volume greatly exceeds the combined
volume of high-density regions.
Indeed, most of the enriched volume is in regions whose density
is between $0.1\bar\rho$ and $\bar\rho$.

\item
As outflows become more anisotropic (smaller opening angle $\alpha$), the
extent of enrichment increases at all densities, low and high. The effect
is most important at low densities, where the efficiency increases
by factors of several. At very-high densities, $\rho=1000\bar\rho$,
regions are already 80\% enriched by volume by isotropic outflows, so there is
little room for an additional increase.

\item
Overall, anisotropic outflows do not strongly favor low-density regions
at the expense high-density ones, unlike what was found in Paper~I.
Anisotropic outflows are more numerous, resulting in higher probability of 
enrichment at all densities. When this effect is factored-out, we find that
the probability that 
a galaxy will enrich systems at densities up to $10\bar\rho$ is
higher for increasingly anisotropic outflows. This results from
evolution of the metal-enriched IGM. Enriched gas tends to be
located relatively near the large-scale structures hosting the
galaxies responsible for this enrichment. As that enriched gas accretes
onto these structures, the effect of anisotropy is partly washed out.

\item

Photoionization suppression of low-mass galaxy formation leads to a number 
of modifications to our enrichment predictions. Since many of these low mass 
galaxies would otherwise
have formed in or near voids of our density distribution, their suppression 
leads to a reduction in the enrichment of low-density regions. Also a 
surprising, 
but not implausible, consequence of this prescription is a fall in volume 
fraction
of enriched regions after $z=3$. This is a consequence of suppression of much 
late galaxy formation combined with the accretion of enriched structures back
onto high-density, low-volume regions.

\end{itemize}

\acknowledgments

This work benefited from stimulating discussions with C\'edric Grenon.
Several of the codes used for analyzing the results of the simulations
have been developed by Paramita Barai and Jo\"el Germain.
All calculations were performed at the {\sl Laboratoire
d'astrophysique num\'erique}, Universit\'e Laval. 
We thank the Canada Research Chair program and NSERC for support. 
MP is supported in part by the Center for Cosmology and Astro-Particle
Physics at Ohio State University.

\appendix

\section{MERGER TREE: DEALING WITH SPLITTERS}

While building merger trees, we discovered that some clusters of ${\rm P^3M}$ 
particles were
actually {\it splitting}, that is, breaking up into two or more components.
Such splitting might be real. A small cluster located in the vicinity of 
a larger one might get stretched by the tidal field to the point where 
it breaks up, or it might get so elongated that the FOF
algorithm identifies it as several clusters. This can also be a purely
numerical effect, which could
occur in two different situations: (1) Two clusters that pass
near each others without merging come briefly into contact. If the FOF 
algorithm
is applied while the clusters are in contact, they will be identified as
one single cluster, that then splits. (2) When two clusters approach each
others on a collision course, they will be identified as one single cluster
as soon as one particle from the first cluster and one particle from the other
cluster
are within a linking length of one another. Then, because of the internal
motion of the particles inside clusters, the two clusters might lose and 
re-establish contact several times during the following timesteps, before
they finally merge for good. These effects are artifacts of the FOF algorithm,
and usually go unnoticed unless one uses a very fine time resolution when 
building merger trees. In this paper, we build cluster catalogs spaced by
time intervals of $2.5\times10^7{\rm years}$, which gives us merger trees with
306 levels between redshift $z=6$ and $z=2$. With such a fine time resolution,
these `splitters' become a real problem, which needs to be addressed. More 
specifically,
since mergers result in starburst, we have to be sure that mergers identified
by the FOF algorithm are real. A merger that immediately follows or
is immediately followed by a splitter might not be real.

\begin{figure}
\includegraphics[width=0.8\columnwidth,angle=90]{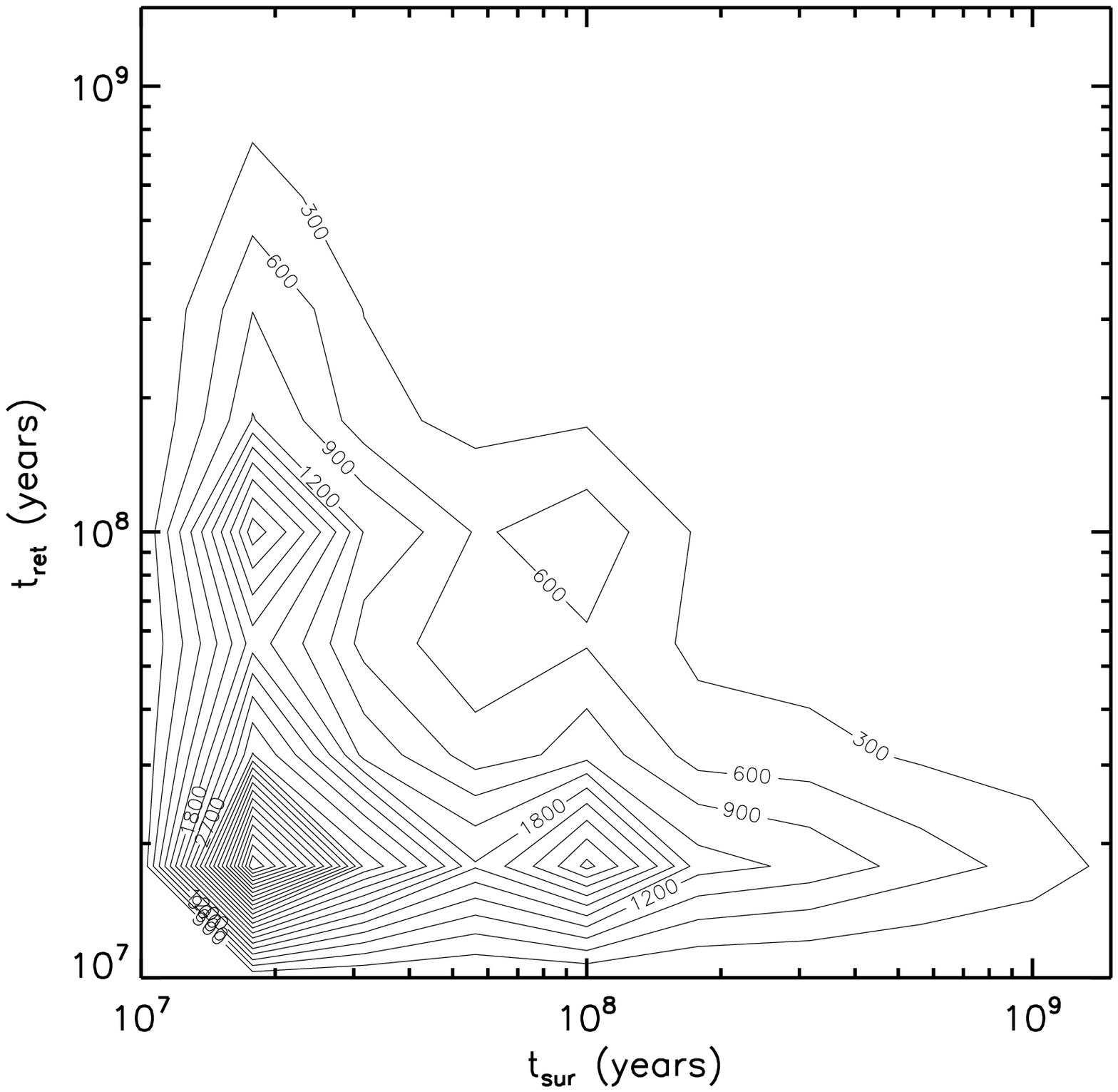}
\caption{Contour plot showing the time $t_{\rm sur}$ that a merger remnant
survives before it splits, vs. the time $t_{\rm ret}$ that a
split lasts before the various parts re-merge together. The contours
show the number of clusters. All clusters have 
either $t_{\rm sur}<t_{\rm stab}$, $t_{\rm ret}<t_{\rm stab}$, or
both, where $t_{\rm stab}=1.5\times10^8{\rm years}$ is the time of stability.
} \label{split}
\end{figure}

We have designed an algorithm that scans our merger trees and eliminate
splitters that appear to be spurious. We start by defining a 
{\it time of stability} $t_{\rm stab}$, which is the minimum time a 
cluster must exist in the simulation to be considered a real cluster. 
When a cluster forms by a merger, and then splits, we calculate the time 
$t_{\rm sur}$ during which the cluster survived before splitting. If
$t_{\rm sur}<t_{\rm stab}$, we consider that the cluster was not present
long enough to be considered real, and therefore the merger that formed that
cluster and the subsequent splitter that destroyed it were both
spurious. This handles the first case described above, that is of
two clusters that briefly come into contact during a non-merging encounter.

This leaves the second case, two clusters that experience a series of 
mergers and splitters before finally merging for good. In this case,
every time the cluster splits, the two pieces will re-merge after a time
$t_{\rm ret}$. If this time is sufficiently short, we assume that the
split and the subsequent re-merging were spurious, and that the cluster 
actually never split. We use the criterion $t_{\rm ret}<t_{\rm stab}$,
where $t_{\rm stab}$ is the time of stability that was used in the other 
criterion.

We have applied this method to the merger tree built from our simulation, 
using a time of stability $t_{\rm stab}=1.5\times10^8{\rm years}$. We
have identified all cases where a cluster splits. For each case, we calculated
the time $t_{\rm sur}$ during which the cluster existed before it split, and
the time $t_{\rm ret}$ elapsed before the various pieces re-merged together
(this can be infinite if the pieces never re-merged).
In Figure~\ref{split}, we show a plot
of $t_{\rm ret}$ vs. $t_{\rm sur}$, where the contours show the actual
number of clusters. Most clusters have either $t_{\rm sur}<t_{\rm stab}$,
$t_{\rm ret}<t_{\rm stab}$, or both. Hence, according to our criterion,
these splitters are not real, and we simply ignore them. A small
fraction of splitters fail both criteria, and are considered
real splitters. Such splitters occur when a small halo is
torn apart by the tidal field of a larger one. This violent process should 
result in a significant reheating of the gas, which can later cool down
and lead to a starburst. Consequently, we treat a halo formed by a
splitter like we treat halos formed by monolithic collapse and
mergers: we calculate its cooling time, and after that time has elapsed,
we start a new outflow. Note that the details of this treatment are not
critical, since real splitters are rare.

\section{THE SPHERICAL COLLAPSE MODEL}

In SB01 and Paper~I, the spherical top-hat collapse model was used to
estimate the size of uncollapsed halos when these halos are hit by 
outflows. The comoving radius of a halo is given by
\begin{equation}
l=\left({2\over9}\right)^{1/3}\!
{R(1-\cos\theta)\over(\theta-\sin\theta)^{2/3}}\,,
\label{rcom}
\end{equation}

\noindent where $R$ is the comoving radius in the 
limit $z\rightarrow\infty$. In equation~(\ref{rcom}),
the parameter $\theta$ is obtained by solving the following
transcendental equation,
\begin{equation}
\left({\theta-\sin\theta\over2\pi}\right)^{2/3}=
{\delta_+(z)\over\delta_+(z_{\rm coll})}\,,
\label{time}
\end{equation}

\noindent where $\delta_+$ is the linear growing mode. As $z$ varies
from $z=\infty$ to $z=z_{\rm coll}$, $\theta$ varies from 0 to
$2\pi$. The physical radius of the halo is given by
\begin{equation}
l_{\rm phys}
=\left({2\over9}\right)^{1/3}\!{R(1-\cos\theta)\over(1+z)
(\theta-\sin\theta)^{2/3}}\,.
\label{rphys}
\end{equation}

\noindent Notice that at high redshift (when $\Omega\approx1$
and $\lambda\approx0$), $\delta_+(z)\approx(1+z)^{-1}$. Hence, the denominator
in equation~(\ref{rphys}) is nearly constant. The time-dependence
in equation~(\ref{rphys}) essentially comes from the factor
$(1-\cos\theta)$ in the numerator. The halo has a physical
size of 0 at $\theta=0$ (corresponding to the big bang), reaches
maximum size at $\theta=\pi$, and collapses to a black hole at
$\theta=2\pi$. 

In practice, halos will not collapse to a black hole. After turnaround,
radial motions will be converted into transverse motions, and eventually
the halo will virialize, with a physical radius that is 1/2 of the
maximum physical radius (that is, the halo shrinks by a factor of
2 in radius after turnaround). We do not have a precise model
for the evolution of the system between turnaround and virialization.
A fair assumption is that the time it takes for the halo to
shrink by a factor of 2 is equal to the time it would have taken
to collapse to a black hole in the absence of virialization
(\citealt{padmanabhan93,do99};
see however \citealt{cl95}). To achieve this result, we modify 
equation~(\ref{rcom}) and (\ref{rphys}), using

\begin{eqnarray}
l&=&\left({2\over9}\right)^{1/3}
{R\chi(\theta)\over(\theta-\sin\theta)^{2/3}}\,,
\label{rcom2}\\
l_{\rm phys}&=&\left({2\over9}\right)^{1/3}\!
{R\chi(\theta)\over(1+z)(\theta-\sin\theta)^{2/3}}\,,
\label{rphys2}
\end{eqnarray}

\noindent where

\begin{equation}
\chi(\theta)=\cases{1-\cos\theta\,,& $\theta\leq\pi$;\cr
\noalign{\medskip}
                    \displaystyle{1\over2}(3-\cos\theta)\,,&$\theta>\pi$.\cr}
\end{equation}

\begin{figure}
\begin{center}
\includegraphics[width=0.8\columnwidth]{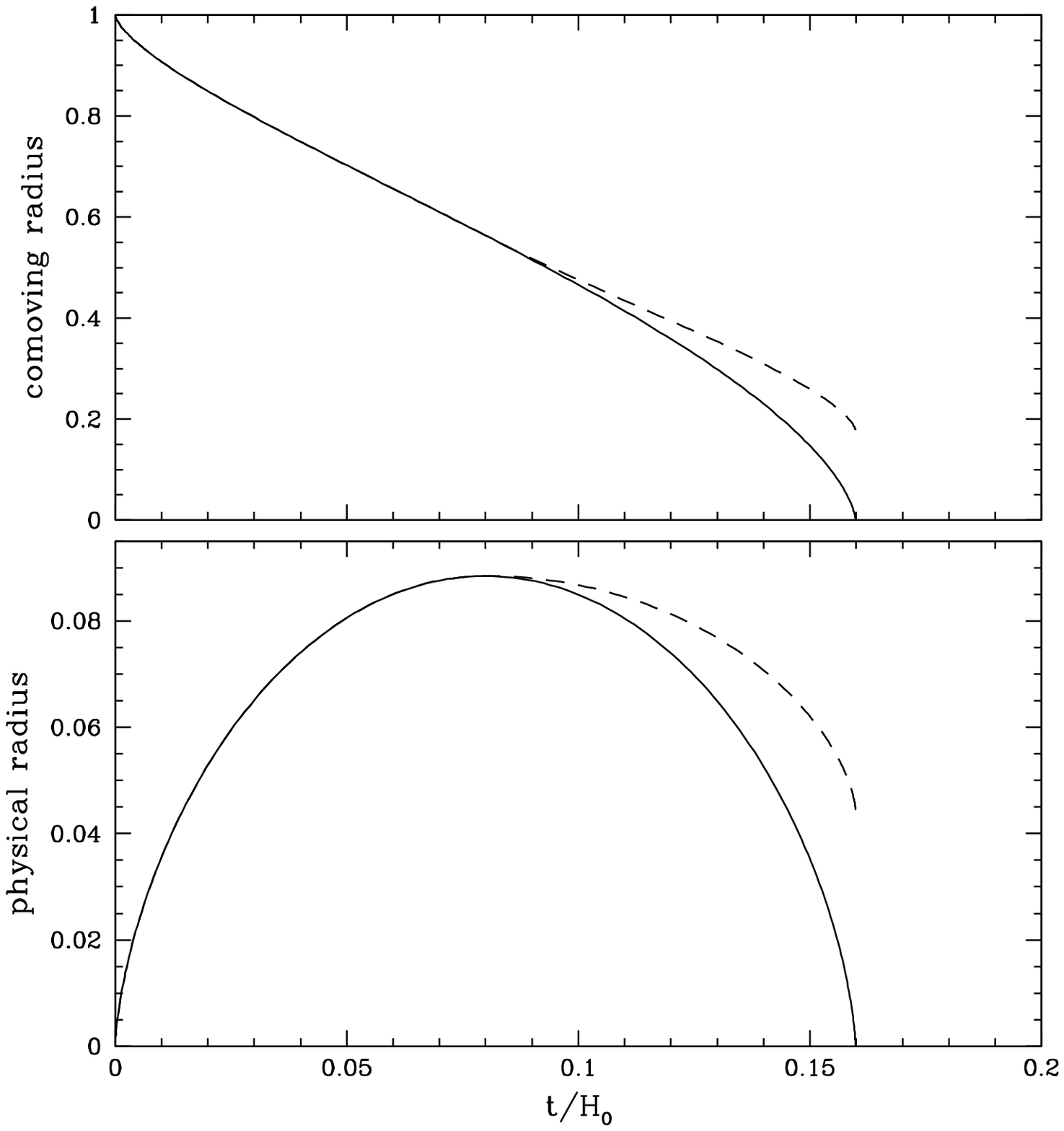}
\caption{Evolution of a spherical halo
collapsing at redshift $z_{\rm coll}=3$ in a universe with $\Omega_0=0.268$
and $\lambda_0=0.732$. The solid and dashed curves show the standard and 
modified spherical collapse model, respectively.
}
\label{spherical}
\end{center}
\end{figure}

Figure~\ref{spherical} shows the evolution of a spherical halo
collapsing at redshift $z_{\rm coll}=3$ in a universe with $\Omega_0=0.268$
and $\lambda_0=0.732$. The comoving radius and physical radius are
plotted versus time in units of the Hubble time $1/H_0$. Since the
evolution is similar for halos of different sizes, we set arbitrarily
$R=1$. The solid curves show the basic spherical model given by
equations~(\ref{rcom}), (\ref{time}) and~(\ref{rphys}). The dashed
curves show the effect of using equation~(\ref{rcom2})
and (\ref{rphys2}). After turnaround,
the halo collapses by a factor close to 2. 

It is convenient to re-express the physical radius in terms of its final value
at $z=z_{\rm coll}$. That value is obtained by setting $z=z_{\rm coll}$ and
$\theta=2\pi$ in equation~(\ref{rphys2}). We get
\begin{equation}
l_{\rm phys}(z)=l_{\rm phys}(z_{\rm coll})
\biggl[{(1+z_{\rm coll})\delta_+(z_{\rm coll})
\over(1+z)\delta_+(z)}\biggr]^{2/3}\chi(\theta)\,.
\end{equation} 

Since at high redshift the quantity in brackets is close to unity,
the time-dependence is essentially given by the function $\chi(\theta)$.

%

\end{document}